\documentclass[11pt]{article}
\usepackage{times}
\usepackage{geometry}
\usepackage[utf8]{inputenc}
\usepackage{enumitem,amssymb}
\usepackage{graphicx}
\usepackage{fancyhdr}
\usepackage{aas_macros}
\usepackage{mdframed} 

\usepackage{hyperref}

\newcommand{\Hi}{H~{\sc i}}
\newcommand{\Hii}{H~{\sc ii}}

\usepackage[authoryear]{natbib}
\setcitestyle{authoryear,open={(},close={)}}

\mdfdefinestyle{theoremstyle}{
innertopmargin=\topskip,}
\mdtheorem[style=theoremstyle]{lrptextbox}{}

\begin{document}

\section{Introduction and history}



The DRAO Synthesis Telescope (ST) is a seven-element east-west aperture synthesis telescope with a maximum baseline of 617 metres, designed initially for \Hi\ imaging in the 21~cm line. A fully sampled u-v plane and small (8.6\,m) antennas make it one of the best telescopes in the world for imaging extended structure. Through incorporation of single-antenna data, ST images represent structure on all scales (also extended to polarization imaging, \citealp{Landecker:Reich:2010}). This White Paper puts forward a plan of renewal which will greatly expand the scientific reach of the telescope, and will provide a flexible testbed for innovations in radio-astronomy technology and software. It will serve as a training ground for the next generation of astronomers and engineers.

In the 1990s, the ST user community came together to propose the CGPS, the major project on the telescope from 1995 to 2007. Forty faculty, postdocs and students in the Canadian astronomical community produced over 200 science papers, 14 Ph.D. and 24 M.Sc. degrees were awarded in Canadian universities, and 27 postdocs worked on the project (12 now hold tenured positions in Canada). The CGPS was funded by NRC and NSERC. The survey data are hosted at the Canadian Astronomy Data Centre (CADC).

 Using the CGPS, Canadian-led teams developed much of our modern understanding of the interplay between multi-phase gas and magnetic fields in the Galactic ISM. Scientific highlights include (a) revealing arcminute structure of Galactic \Hi; (b) tracing the Galactic magnetic field from Rotation Measures (RMs) of extragalactic sources seen through the disk, establishing the ``RM grid'' technique (\citealt{Brown:Taylor:2001,Brown:Taylor:2003}); (c) the detection of widespread cold \Hi, the first stage of star formation, seen absorbing emission of more distant, warmer \Hi\ \citep{Gibson:2005}; (d) the detection, analysis and interpretation of diffuse polarized emission from the Galactic plane \citep{Landecker:Reich:2010}; and (e) an improved understanding of the rotation curve of the outer Galaxy, leading to better kinematic distances \citep{Foster:MacWilliams:2006}. The DRAO Planck Deep Fields (DPDF) project (NSERC funded) subsequently used the ST (2006 to 2009) to produce ultra-deep images of two substantial fields. Highlights include (a) deep polarization imaging to study the faint population of compact extragalactic sources \citep{Grant:Taylor:2010}; and (b) studies of foreground dust physics by linking \Hi\ and infrared data.

The telescope continues to operate as a fully subscribed National Facility; since LRP2010 the ST has made observations for 49 astronomers (34 were Canadian). These were based on 26 proposals from 27 institutions (10 Canadian) in 11 countries. Twelve graduate and undergraduate students have been involved. Research highlights since the completion of the CGPS include (a) detection of extensive correlation between polarization structures and high-velocity clouds (HVCs), giving insights into the magnetic field in the Galactic halo;(b) extended rotation curves for nearby galaxies;(c) follow-up observations of FRBs detected by CHIME to measure foreground contributions to dispersion measure.


\section{Renewal}

Three roles are envisaged for the renewed ST: it will be a forefront scientific instrument for new challenges, a testbed for new radio astronomy technologies, and a training ground for students in astronomy and in engineering. These goals fit well with the role of DRAO as a National Facility. There has already been substantial progress on renewal, driven by university partners and their graduate students. The concept of telescope-as-a-testbed, combining scientific and engineering roles, has support at the highest levels of the Herzberg Astronomy and Astrophysics (HAA). HAA develops advanced radio astronomy technologies: uptake of those technologies by the astronomical world succeeds when new concepts can be shown to work doing forefront science. 

\subsection{Progress to date on renewal}

\subsubsection{Feed antennas}

Research on wideband single-pixel feeds began in 2012, in collaboration with Professor Thomas Johnson (Engineering, UBC Okanagan). M.Sc.\ student Andr\'e Johnson designed and tested a feed operating over the entire band 1 to 8 GHz, based on new concepts. Ph.D. student Xuan Du (also UBCO) has now perfected a design covering 400 to 1800 MHz, based on the same concept. Two coaxial waveguides, one inside the other, cover 400 to 800 MHz and 900 to 1800 MHz (800 to 900 MHz is dominated by cell phone signals). Prototypes will be tested in 2019. Compared to competing wideband technologies 
the design offers compact size, light weight, and ease of manufacture. In bandwidth it out-performs the classic corrugated horn designs.

\subsubsection{GPU-based correlator}
\label{gpu-correl}

A new correlator for the telescope has been developed by M.Sc. student Pamela Freeman (Physics and Astronomy, University of Calgary, supervisor Jo-Anne Brown), with involvement of Keith Vanderlinde and Andr\'e Renard (University of Toronto). Based on technology developed by UofT for CHIME, the correlator has a bandwidth of 400 MHz, $\sim$10 times the bandwidth of the existing machine. The new correlator is now under test on the telescope. (A new proposal for collaboration of these partners on further development of this correlator has been submitted to NRC for funding, with results expected in December 2019).

\subsubsection{Low-noise amplifiers}
 
Professor Leonid Belostotski of the University of Calgary (Engineering) has a long-standing collaboration with DRAO. His group is a world leader in the design of low-noise amplifiers (LNAs) for radio astronomy. They were the first to use CMOS technology in LNA design, and their LNAs have been tested on the existing ST for several years \citep{belo14}. For ST renewal Thisara Katanunga (M.Sc. 2018) designed an LNA for the 400 to 800 MHz band with excellent performance. Ph.D. student Alexander Sheldon is developing a CMOS amplifier for the 900 to 1800 MHz band; it already has very good performance operating at room temperature, but cooling with small cryogenic refrigerators will be part of his work. The renewed ST as a testbed will be a valuable asset for the U. Calgary group as they pursue other scientific and commercial applications of their research.



\subsection{A new proposal: {\em Rethinking Astronomical Interferometry}}

An ambitious proposal, titled as above, has been submitted to an internal NRC competition. The proposal will equip the ST to test new ideas that have the potential to profoundly change the technology of interferometry, with benefits far beyond the ST, carrying radio astronomers beyond today's technological and cost limitations. 

 The requested sum is \$2 million. Over 2020 -- 2023 this will equip 4 of the 7 ST antennas with the feeds and LNAs described above, with a highly innovative digitizing system, and with a prototype of the correlator designed at DRAO for the SKA. The proposed work will test new ideas of profound significance to the future of interferometry in radio and optical astronomy. In the following sections we describe these innovations and their potential for major payback to the Canadian and international astronomical communities.

 The results of the NRC competition will be known in November 2019. To fit the proposal into the available funding the number of antennas had to be limited to 4. With the NRC decision known, other funding sources (e.g. CFI) will be investigated so that all 7 antennas can be refitted. The estimated requirement is a further \$0.5 million.


\subsubsection{Incoherent clocking of interferometer elements}

Radio astronomy interferometers built in the past used complex and costly systems to maintain coherence of the elements by phase-locking local oscillator signals or sampling clocks. The proposal here is to digitize at each antenna using individual sampling clocks with no attempt at coherence. At each antenna the digitized signal contains all the information received by that antenna. That information and the sampling clock signal  are both returned to the centre of the telescope, and coherence among the interferometer elements can be re-established prior to correlation \citep{carl18}. The advantage of this system is that it can use standard commercial (low-cost) grades of optical fibre as the transmission medium. Furthermore, this technique overcomes cost and technology barriers to establishing coherence over very long baselines. It has potential application to the extension of ALMA baselines to 300 km, to the ngVLA, and possibly to the SKA. It may be possible to use commercial packet-switched fibre networks; it could then be extended to trans-continental baselines.

Using NRC funds (from an earlier funding competition), DRAO now has a prototype system and is equipped to test incoherent clocking in the laboratory over fibre lengths up to 360 km in 10 km increments (the fibre will be above ground, on poles, in conditions resembling a commercial communications network).  Laboratory tests can go only so far: to go further demands application in a working telescope. A several-day test between the Jansky VLA and the Pie Town station of the VLBA (using a 105-km fibre link) is scheduled for 2020. Realizing the huge potential of incoherent clocking requires protracted on-sky testing, and that will be provided by the renewed ST.

Under the proposal submitted to NRC the incoherent clocking concept will be tested in {\em{optical}} interferometry (by a partner group within NRC). Success will mean that that techniques long established in radio astronomical interferometry, digital delay and digital correlation, can now be applied to optical interferometry. To achieve astronomically meaningful bandwidths requires digital processing of signals with bandwidths of hundreds of GHz, so potential applications are well in the future. Nevertheless, the opportunity to push this frontier is now with us. 

\subsubsection{Averaged clock oscillators}

Hydrogen maser clocks provide synchronization in VLBI today. The current proposal for the ngVLA sees a maser at each of 80 distant antenna sites. With incoherent clocking, masers may no longer be needed. Spectral purity is still required of each clock, but the exact frequencies become unimportant. On the renewed Synthesis Telescope we will test a low-cost scheme to generate a clock signal of great spectral purity using commercial crystal oscillators and coherent averaging methods. Substituting such devices for masers on the ngVLA could increase the coherence time of the outer antennas to that of the atmosphere, simplifying data reduction (and sharply reducing cost).

\subsubsection{Correlator}

The DRAO digital engineering team, in collaboration with Canadian industry, has developed the TALON correlator/beamformer for SKA1-Mid, which has been adopted as the reference design for both SKA1-Mid and the ngVLA. A working prototype board exists at DRAO. A version of this correlator (obviously much smaller than the SKA or ngVLA will require!) will be built for the renewed ST. 

The TALON correlator uses advanced digital signal processing techniques to produce extremely high-fidelity visibilities
(with flexible spectral resolution down to 200 Hz) even in strong RFI environments. The correlator uses the latest technology, a 14 nanometre silicon Field Programmable Gate Array (FPGA) and fibre-optics signal paths. When provided with frequency and time signals from a commercial GPS receiver, the correlator will translate the incoherently clocked inputs to a common frequency before correlation and produce time-stamped visibilities that can then be processed to images with the Common Astronomy Software Applications (CASA) package. All correlator functions needed to carry out the science program described in this document are already present in the TALON system. For the renewed ST the correlator will occupy no more than one 19-inch equipment rack.

The value of testing this correlator under real on-the-sky conditions cannot be overstated. Laboratory tests can uncover a lot of problems, but application in a telescope goes deeper.
The rigorous testing that this correlator will receive in front-line astronomical applications will greatly strengthen the case for adoption of the Canadian design for the SKA, for the ngVLA, and potentially even for ALMA. The construction of the full-scale correlator is then much more likely to fall to Canadian industry, with benefits to Canadian astronomy, and to Canadian society. 

\subsection{Outlook: The further future}

\subsubsection{Incorporating other antennas} \label{sec:antennas}

The seven antennas of the present Synthesis Telescope are capable of operating up to 2 GHz. There will inevitably be a demand to use the testbed to try out equipment and ideas for higher frequencies, and the renewed telescope will be designed to permit incorporation of further antennas. The goal is to have a total of ten antennas, with seven always available for science and up to three available for technology testing.

DRAO has on site a composite antenna, an offset Gregorian design of aperture 15 m, with a 50 GHz surface. This carbon-fibre reflector was built by the DRAO Composite Antenna group. DRAO also has a 10-m composite antenna capable of 10 GHz operation, as well as an offset 15-m reflector capable of 20 GHz operation (but no mount for that reflector). The renewed ST will be designed so that some or all of these antennas can be incorporated as the need arises. Sub-arrays will be possible so that science observations and testbed observations can occur concurrently, and it will also be possible to use the high-frequency antennas in stand-alone mode.
 
 Once again, the value of on-the-sky-tests in an operating telescope cannot be overstated. The DRAO Composite Antenna group has provided an 18-m design, which has become the reference antenna design for the ngVLA. Testing composite antennas within the ST testbed will advance prospects for the adoption of DRAO technology in astronomy projects around the world.
 
\subsubsection{Future correlator developments}

Graphical Processing Unit (GPU) technology is the foundation of the innovative correlator developed by the University of Toronto group led by Keith Vanderlinde. That group is now investigating other radio astronomy applications, among them the new correlator now being tested on the ST by the University of Calgary (see Section~\ref{gpu-correl}). The renewed ST will provide opportunities to demonstrate the new designs they are now developing. Two correlators can run in parallel, testing the second design against the established performance of the first.

\subsubsection{An open testbed for Canadian innovation}
Developing cutting-edge instrumentation and exploring novel technologies requires on-the-sky testing, but traditionally this has relied on individually negotiated access to facilities which are used to support scientific observers, but not really set up to deal with outside engineers or test equipment.  This has worked reasonably well for single-dish telescopes, but most modern radio instruments are interferometers, and no open-access testbed interferometer exists. The ST renewal aims to fill this gap, providing opportunities for technical development and engineering students that observatories have traditionally offered only for scientific exploration and astronomical training. This will be accomplished through a combination of well-defined (and documented) interfaces, access to the experienced and diverse HAA technical and scientific staff, and ensuring that the renewal program builds this philosophy its design from the very start.  The vision is to offer technical innovators across Canada access to a small, well-understood interferometer, for substantial periods of time, at multiple points along the signal path.

 \section{Science with the renewed Synthesis Telescope}

There are a number of possible scientific endeavours that might be undertaken with the renewed Synthesis Telescope. This discussion is based on the telescope parameters listed in Table~\ref{specs}, with emphasis placed on transformational science. The list is not exhaustive, and is not prioritized: the user community, which will grow in response to the new capabilities of the telescope, will steer future science directions. The renewed ST will have access to the \Hi\ line, the four OH transitions, and radio recombination lines (RRLs). These topics are treated in separate sections in the following discussion, but the intention is that all of these lines can be measured commensally.

\begin{table}[!htbp]
\caption{The New DRAO Synthesis Telescope}
\label{specs}
\centerline{
\begin{tabular}{ll}
\hline
Number of antennas           & 7 \\
Antenna diameter             & $8.6${\thinspace}m \\
Frequency coverage           & 400 to 800\,MHz and 800 to 1800\,MHz simultaneously \\
Feeds                        & Wideband single-pixel \\
System temperature           & 25\,K plus sky \\
Minimum baseline              & $\sim$12.9\,m \\
Maximum baseline             & $\sim$617\,m \\
Instantaneous continuum bandwidth & 1300\,MHz \\
Spectral-line zoom channels  & 15k per 200~MHz subband \\
Angular resolution, 400\,MHz & 2.8 arcminutes \\
Angular resolution, 1800\,MHz & 40 arcseconds \\
\hline
\end{tabular}}
\end{table}

Through work with the Synthesis Telescope, DRAO has developed exceptional scientific and technical skills. These will be exploited in the renewal project, and include the following: {\bf{(i)}} The ST is currently the world's leading telescope for imaging polarized signals over wide fields;
{\bf{(ii)}} Techniques have been developed for incorporating single-antenna data with aperture-synthesis data to ensure that all structure, from the largest to the resolution limit, is truly represented in images. These techniques are relevant to total-intensity data, polarization data, and spectral line data;
{\bf{(iii)}} Single-antenna polarization data are available to DRAO. The Global Magneto-Ionic Medium Survey (GMIMS -- led from DRAO -- see White Paper E064:Cosmic Magnetism), will soon publish data for the entire Northern sky from 1280 to 1750 MHz. The CHIME project will produce a high-precision map of the Galactic polarized emission from 400 to 800 MHz on the timescale of a few years;
{\bf{(iv)}} All-sky \Hi\ data are available from the Effelsberg EBHIS survey \citep{Winkel16};
{\bf{(v)}} DRAO has successfully developed techniques for measuring and correcting for
instrumental polarization across the field of view of synthesis images, enabling precision wide-field polarization imaging;
{\bf{(vi)}} DRAO has well-developed software for correcting for Faraday rotation in the ionosphere. This is essential for high-fidelity polarimetry at all frequencies below 1 GHz, and is desirable for polarimetry at frequencies as high as 1.5 GHz. This software is based on public data from GPS clocks spread across the world. DRAO is well placed, with a precision GPS clock on site and a large number of GPS stations within a few hundred kilometres;
{\bf{(vii)}} DRAO has outstanding skills in processing aperture-synthesis data from small antennas.

\section{Scientific opportunities}
\label{focus}

\subsection{Wideband polarimetry of Galactic emission with arcminute resolution}
\label{Faraday}

A transformative science role for the renewed ST is polarimetry of the extended Galactic emission. Among telescopes capable of exploring this subject, the ST will be unique in terms of wavelength coverage and angular resolution, making this the dominant telescope for detailed study of the Galactic Magneto-Ionic Medium (MIM). Single antenna data suitable for incorporation with ST data will be available from GMIMS so all structure from the largest to the few-arcminute resolution limit of the telescope can be accurately portrayed. The frequency range of the renewed ST, 400 to 1800\,MHz, closely matches that of GMIMS. This is a crucial frequency range in terms of depolarization and Faraday depth coverage: at lower frequencies Faraday rotation so dominates that only quite local phenomena can be probed. At higher frequencies Faraday rotation is so weak that huge bandwidths are required.


The Faraday depth spectrum is produced by applying Rotation Measure Synthesis to wideband polarization data (Stokes $Q$ and $U$ measured in many frequency channels). The resolution in Faraday depth depends mostly on the longest wavelength of the data and the maximum width of the Faraday depth structures that can be successfully mapped depends mostly on the shortest wavelength. In other words, wavelength coverage is the key to successful investigation of the MIM. With the telescope parameters listed in Table 1, the resolution in Faraday depth will be 7.1\,${\rm{rad}}\thinspace{\rm{m}}^{-2}$ and the largest detectable Faraday depth feature will 112\,${\rm{rad}}\thinspace{\rm{m}}^{-2}$. When seen together with the exceptional ability to image large structure and to incorporate single-antenna data, the renewed ST stands out from the competition for mapping the Galactic diffuse emission.


The magnetic field, the gas, and the cosmic rays dominate the energy budget of the ISM in almost equal shares (Ferriere 2001). We have detailed three-dimensional knowledge of the gas distribution and motions, through \Hi\,and molecular-line data, all with arcminute resolution. Through Faraday spectra, GMIMS is starting to provide the third dimension for magnetic field studies, but only on one-degree scales. There are some valuable investigations from the WSRT, from LOFAR, and from the MWA, but nothing matching what the renewed ST could do. 

The ST can reach mostly the outer Galaxy. Is this enough? Experience from the CGPS suggests strongly that the benefit of the outer Galaxy is its relative simplicity. Towards the outer Galaxy we see primarily the Perseus arm. The Perseus arm is a wonderful laboratory for the study of all the things that make the ISM tick. Recent experience suggests that observations at higher latitudes might have beneficial simplicity over observations in the disk.

One should not study the polarized emission alone. Polarized intensity represents a convolution of synchrotron emission (born of cosmic ray electrons and magnetic fields) with Faraday rotation (which traces the magnetic field and ionized gas) and thus is influenced by all three important constituents of the ISM. We cannot understand one ISM component in isolation. The renewed ST will be able to map the \Hi\,emission and RRLs simultaneously to trace the neutral and ionized gas in addition to the diffuse polarized emission.

With arcminute resolution we will be able to investigate the role of magnetic fields in interstellar processes of all kinds. What is the intensity and structure of the magnetic field in stellar wind bubbles? Is the magnetic field enhanced in regions where cold \Hi\,is seen in self-absorption (presumably the first stages of star formation)? Can we better understand the magnetized turbulence that generates so much small-scale structure in polarization images? Arcminute resolution provides parsec physical resolution in the closest spiral arms, and significant processes in the Galactic ecosystem will only be understood once that scale is reached.

\subsection{Rotation measures of extragalactic sources as probes of the foreground Magneto-Ionic Medium}

Aside from the intrinsic interest of the Faraday structure of extragalactic sources, the Rotation Measures determined using these sources provide useful probes of foreground magneto-ionic material. One of the major successes of the CGPS came from mapping of the Galactic magnetic field through such probing \citep{Brown:2010}. The existing telescope is able to provide one to two RMs per square degree depending on location. The new telescope, through its superior sensitivity will provide five to ten. This will allow a more detailed probing of the large-scale field.

\subsection{Wideband Faraday rotation studies of extragalactic sources}
\label{wideband-eg-sources}

Wideband observations of extragalactic sources show complex Faraday structure intrinsic to the sources \citep{OSullivan:Brown:2012}. This Faraday complexity, not evident from older, narrow-band observations, reflects emission mechanisms and magnetic field configurations internal to the source. The above paper was based to 1--3 GHz observations; similar searches for Faraday complexity would be very successful with a telescope capable of observing down to 400~MHz, (with the caution that some sources may no longer be polarized below 1\,GHz). Data from the Synthesis Telescope covering 400 to 1800\,MHz will be complementary to the VLASS survey, covering 2 to 4\,GHz; together the two datsets will cover a very large range of $\lambda^2$.

\subsection{\Hi\ Absorption of Polarized Galactic Emission}
\label{HI-absorption}

\Hi\ absorption has been a powerful tool, providing kinematic distances for Galactic objects. The target source must produce a signal in the telescope of $\sim$15\,K to avoid confusion of genuine absorption with \Hi\ self-absorption or faint emission in a reference patch of sky. These problems can be circumvented by searching for absorption of {\em{polarized}} emission because \Hi\ emission is totally unpolarized. \citet{Kothes:2004} demonstrated this technique using the VLA to determine distances for two faint supernova remnants (SNRs). With the renewed ST it will be possible to extend \Hi\ absorption measurements to about 40 SNRs for which no effective distance determination has been made (as long as they have detectable polarized emission at 1420~MHz). Accurate distances to these SNRs will open the door to credible analyses of their physical properties.

This technique was first demonstrated by \citet{Dickey:1997} who used it to determine minimum distances to a few features of the extended Galactic polarized emission. No further work on this topic has been done. We now propose to apply the technique to the extended polarized emission using the renewed ST. The exceptional ability of the ST to image extended emission can provide results that no other telescope can match. Studies of the distribution of polarized emission will then have three-dimensional data where previously the only data were maps at isolated single frequencies. \Hi\ absorption of polarized emission, together with Faraday depth studies (Section~\ref{Faraday}), can provide transformational data on the distribution and strength of Galactic magnetic fields.

Graduate student Rebecca Booth (Physics \& Astronomy, University of Calgary, co-supervisors Jo-Anne Brown and Tom Landecker) will make test observations in 2020, exploring \Hi\ absorption of polarized emission from SNRs using the present ST. This is a first step towards a significant science program with the renewed telescope.

\subsection{\Hi\ surveys}

\Hi\ data will be available with every observation. Rather than attempt to list all possible uses of \Hi\ imaging, we concentrate on one application, informed by a recent spectacular discovery.

In mapping a region at medium Galactic latitudes containing a well-known group of high-velocity clouds (HVCs) remarkable correlations between \Hi\ gas and linearly polarized radio emission have been found (Forster et al, Nature, submitted). This astounding symmetry suggests that HVCs are compressing and altering large-scale magnetic fields in the Galactic halo as they rain down onto the Galactic disk. HVCs may be direct probes of the halo magnetic field through the imprint they leave on linearly polarized radio emission. Subsequent examination of the ST archive has revealed many more similar correlations. The renewed ST will obtain Faraday spectra for these polarization features and there is a real prospect of measuring, and even mapping, the magnetic field in the halo.

\subsection{OH surveys}

The renewed ST will {\em{not}} have the sensitivity to map thermal OH, but it is certainly competitive in blind searches for OH masers in the Galactic plane, with sensitivity $\sim$10 mJy/beam per spectrometer channel. Such data would be complementary to the THOR survey \citep{Beuther:Bihr:2016} made with the Jansky VLA and SPLASH \citep{Dawson:Walsh:2016} made with the Parkes Telescope. The TALON correlator will have the flexibility to bring zoom modes to the four $\lambda=18\,{\rm{cm}}$ OH transitions. The telescope could also measure magnetic fields in maser sources through Zeeman splitting, complementing the southern MAGMO survey \citep{Green:McClure-Griffiths:2014}.

Superradiance, a brief peak in intensity, has been observed in OH transitions at 1612\,MHz and 6.7 GHz \citep{Rajabi:Houde:2019}. The ST, with its slow observing cadence, typically 12 days, might discover such events.

\subsection{Radio Recombination Lines}
\label{recomb}


The significance of the Warm Ionized Medium (WIM) is discussed in White Paper E081 (Interstellar Medium). A truly comprehensive view of the interstellar gas demands detailed mapping of the WIM, and the renewed ST can make a major contribution here by mapping in radio recombination lines (RRLs). Optical H$\alpha$ surveys such as WHAM \citep{Haffner:Reynolds:2003} offer the greatest sensitivity in the local ISM but are limited by extinction in many directions, and the $1^{\circ}$ WHAM beam becomes ineffective at large distances. RRL emission is very faint, but adequate sensitivity can be achieved by stacking signals from many transitions. Existing surveys using this method are THOR \citep{Beuther:Bihr:2016} made with the JVLA, SIGGMA \citep{Liu:Anderson:2019} made with Arecibo, and GDIGS \citep{Luisi:Anderson:2018} made with the GBT. These surveys have concentrated on the inner Galaxy.
 
 Stacking many RRL transitions is simple in the TALON correlator. The lines are broad, and the velocity resolution required is about 10 km\,s$^{-1}$. 89 Hn$\alpha$ lines are accessible in the frequency range of the renewed ST. Assuming that 50 of them are in RFI-free parts of the band, the sensitivity achieved can be 13 mK or 560\,cm\,pc$^{-6}$ for a 3-$\sigma$ detection. This is $\sim$3 times the sensitivity of the THOR survey and roughly the same sensitivity as the SIGGMA survey, currently the most sensitive survey available (but limited to the two narrow strips where Arecibo can see the Galactic plane). Angular resolution will be a few times inferior to THOR but $\sim$2 times better than SIGGMA and comparable to GDIGS, with the ability to image anywhere in the Northern sky.

There are many areas of ``big-picture'' RRL science that could be explored with an upgraded ST. For example, line and continuum temperature measurements can be turned into emission measure and electron temperatures, and velocities turned into kinematic distances: one could directly measure the electron temperature gradient with distance from Galactic centre. Another area is extinction: the ratio of Hn$\alpha$ radio lines to optical H$\alpha$ gives the total extinction $A_V$, and one could map $A_V$ across the sky at various kinematic distances. RRLs at arcminute-resolution will offer unprecedented views of the distribution of helium-4 and carbon-12 in \Hii\ regions. 

\section{Operating costs}

Operating costs for the ST have been low. Since 2010 the telescope has been operated with 2 FTEs of DRAO staff, one scientist and one technologist. In addition, one scientist spends time helping telescope users, particularly students, but some of that time is investment in that scientist's collaborations. The workforce in the future will be no larger, except when new technologies are being installed on the telescope for test. The cash cost of maintaining the telescope is of the order of \$20,000 per year, a trivial addition to the infrastructure cost of operating DRAO. Future maintenance costs should be similar. This excludes the cost of future tests of new equipment and techniques; those will have to be borne by the budgets for the development of that equipment. 

\section{Training opportunities}

DRAO has a long track record of student participation in research. This has included undergraduates and graduate students, in science and in engineering. DRAO is a unique institution where astronomers and engineers collaborate readily, creating an excellent training environment. The bulk of the development work for the present ST was done by Canadian graduate students in engineering schools. Development work for the renewal is following the same pattern, with five graduate students currently developing hardware for the ST, along with new observing strategies.

In the age the SKA and remote observing, students will likely have no need (or ability) to go to an observatory. Consequently, the divide between those that build and those that observe will only grow. By contrast, the ST is close and accessible. 
 While it is not a `big' telescope, the ST offers students the opportunity to have hands-on experience with both equipment and science projects. This combination lends itself nicely to the development of an NSERC CREATE training program, which will be the next goal after securing renewal funding for the ST.  
 
 \vspace{4mm}

\begin{lrptextbox}[How does the proposed initiative result in fundamental or transformational advances in our understanding of the Universe?]

The renewed ST will make transformational advances in our understanding of the role of magnetic fields in interstellar processes in the Galaxy. It will be the best telescope in the world to do this. Understanding of the Galactic Warm Ionized Medium will be advanced through observations of Radio Recombination Lines.

\end{lrptextbox}

\begin{lrptextbox}[What are the main scientific risks and how will they be mitigated?]

The science initiatives proposed here depend on broadband observations, well outside the small radio astronomy spectrum allocations. The DRAO site is protected from radio frequency interference (RFI) by municipal, provincial, and federal regulations. It is a national asset, and the astronomical community should work with governments to ensure that its protection is maintained. A decline in quality of the RFI environment will lead to a loss of scientific opportunity. This risk is partly mitigated by development of increasingly clever RFI recognition algorithms. Active research in this field occurs at DRAO and DRAO's university partners. 

\end{lrptextbox}

\begin{lrptextbox}[Is there the expectation of and capacity for Canadian scientific, technical or strategic leadership?] 
Canada has a strong community of ISM scientists. We confidently expect the Canadian user community to grow in response to the enhanced opportunities offered by the renewed ST. In its role as an open-access prototyping testbed interferometer, the renewed ST will be a unique asset for Canada, and will allow testing transformational ideas quickly.  This encourages the blue-sky thinking that leads to great innovations.

\end{lrptextbox}

\begin{lrptextbox}[Is there support from, involvement from, and coordination within the relevant Canadian community and more broadly?] 

Canadian universities are already heavily invested in engineering aspects of the renewal. This White Paper is a collaborative work of DRAO and university colleagues. The present ST has a community of users in Canada, and the science thrust of this proposal is aligned with their research interests.  The technical ideas targeted for initial prototyping have the potential to revolutionize radio design, and are being closely watched by all the major next-generation observatories.
The success of CHIME shows the growing strength of radio telescope engineering in Canada. The interest in prototyping HIRAX and CHORD dishes at DRAO and the use of the well-understood Galt 26m for CHIME calibration, together with current technological exploration using the ST, demonstrates community interest in a more approachable prototyping and training interferometer.

\end{lrptextbox}

\begin{lrptextbox}[Will this program position Canadian astronomy for future opportunities and returns in 2020-2030 or beyond 2030?] 

Radio astronomy is a science driven by technical innovation. Canada needs a strong technical presence on the international stage in order to influence the scientific directions of the projects in which we choose to invest.The Jansky VLA is a good example - Canada built the cutting-edge correlator that makes this such a powerful telescope, and that opened the door to ALMA for us. The renewed Synthesis Telescope will provide the test bed for new technologies and the training ground for the telescope builders of the next generation. The technology that will be on the telescope in the first instance has global implications for astronomy, with potential applications to ALMA, SKA, and ngVLA. The renewal is already providing skill development in Canadian universities. In this era of large international projects it is hard to find training opportunities for graduate students in astronomy that have a hands-on component. These are benefits that are very difficult to obtain elsewhere. 

\end{lrptextbox}

\begin{lrptextbox}[In what ways is the cost-benefit ratio, including existing investments and future operating costs, favourable?] 

The renewal of the ST builds on the infrastructure of the existing telescope. The proposed renewal is a small-budget item that brings large rewards. Operating costs are very low.

\end{lrptextbox}

\begin{lrptextbox}[What are the main programmatic risks
and how will they be mitigated?] 

Key aspects of the renewal of the telescope depend on NRC funding. The new instrumentation ideas described here have been put forward to an internal NRC competition where 14 projects were submitted and only 2 will be funded. The NRC proposal, when funded, will support all three objectives, science return, technology testbed, and training facility. A reduced budget could provide much of the scientific {\em{or}} the engineering capability, but not both. In the event that NRC rejects the proposal, the project will be sub-divided and funding sought through other NRC competitions and through CFI. Support for the training aspects will be requested through the NSERC CREATE program. On the positive side, some key technology developments (feed antennas, LNAs) have already been successful, with NRC and NSERC funding. Incoherent clocking has already won some NRC funding in an earlier competitive process. The LRP should express strong support for the renewal of the ST to support the telescope in all three roles, as a scientific instrument, an engineering testbed, and a training facility. 

\end{lrptextbox}

\begin{lrptextbox}[Does the proposed initiative offer specific tangible benefits to Canadians, including but not limited to interdisciplinary research, industry opportunities, HQP training,
EDI,
outreach or education?] 

The renewed ST will provide a testbed for technological innovation strongly relevant to major radio astronomy initiatives, SKA, ALMA and the ngVLA. Canada has investment and scientific interest in all three. By demonstrating the TALON correlator in a front-
line telescope this project will advance the prospect that the Canadian design is adopted and is subsequently built in Canadian industry. The renewal is already training Canadian graduate students in science and engineering departments, and will continue to do so, developing the future generation of telescope builders and astronomers. Without any special effort this work is already attracting women graduate students. Of the graduate students whose work is described in this White Paper, 3 are men and 2 are women.

\end{lrptextbox}

\section{Recommendation}

The LRP Panel should recommend funding the renewal of the DRAO Synthesis Telescope. At a cost of approximately \$3M, the renewal will provide transformational science opportunities in Cosmic Magnetism and Interstellar Medium science, significant fields with strong Canadian leadership. The telescope will also be a testbed for new technologies with high potential for benefit to international projects in which Canada has a stake. Both the National Research Council and universities are using the telescope to test novel ideas, and the renewal of the telescope is already providing training for graduate students. When operating, the renewed telescope will offer excellent training opportunities within Canada, hands-on experience that cannot be obtained from large international projects.

\setlength{\bibsep}{0.0pt}

\bibliography{output_references} 

\begin{thebibliography}{}
\expandafter\ifx\csname natexlab\endcsname\relax\def\natexlab#1{#1}\fi

\bibitem[{{Belostotski} {et~al.}(2014){Belostotski}, {Haslett}, {Veidt},
  {Landecker}, {Gray}, {Hovey}, {Sheehan}, \& {Messing}}]{belo14}
{Belostotski}, L., {Haslett}, J.~W., {Veidt}, B.~G., {et~al.} 2014, in 16th
  International Symposium on Antenna Technology and Applied Electromagnetics
  (ANTEM), Waterloo, Ontario

\bibitem[{{Beuther} {et~al.}(2016){Beuther}, {Bihr}, {Rugel}, {Johnston},
  {Wang}, {Walter}, {Brunthaler}, {Walsh}, {Ott}, {Stil}, {Henning},
  {Schierhuber}, {Kainulainen}, {Heyer}, {Goldsmith}, {Anderson}, {Longmore},
  {Klessen}, {Glover}, {Urquhart}, {Plume}, {Ragan}, {Schneider},
  {McClure-Griffiths}, {Menten}, {Smith}, {Roy}, {Shanahan}, {Nguyen-Luong}, \&
  {Bigiel}}]{Beuther:Bihr:2016}
{Beuther}, H., {Bihr}, S., {Rugel}, M., {et~al.} 2016, \aap, 595, A32

\bibitem[{{Brown}(2010)}]{Brown:2010}
{Brown}, J.~C. 2010, in Astronomical Society of the Pacific Conference Series,
  Vol. 438, The Dynamic Interstellar Medium: A Celebration of the Canadian
  Galactic Plane Survey, ed. R.~{Kothes}, T.~L. {Landecker}, \& A.~G. {Willis},
  216

\bibitem[{{Brown} \& {Taylor}(2001)}]{Brown:Taylor:2001}
{Brown}, J.~C., \& {Taylor}, A.~R. 2001, \apjl, 563, L31

\bibitem[{{Brown} {et~al.}(2003){Brown}, {Taylor}, \&
  {Jackel}}]{Brown:Taylor:2003}
{Brown}, J.~C., {Taylor}, A.~R., \& {Jackel}, B.~J. 2003, \apjs, 145, 213

\bibitem[{Carlson(2018)}]{carl18}
Carlson, B.~R. 2018, Electronics Letters, 54, 909

\bibitem[{{Dawson} \& {Walsh}(2016)}]{Dawson:Walsh:2016}
{Dawson}, J.~R., \& {Walsh}, A.~J. 2016, in IAU Symposium, Vol. 315, From
  Interstellar Clouds to Star-Forming Galaxies: Universal Processes?, ed.
  P.~{Jablonka}, P.~{Andr{\'e}}, \& F.~{van der Tak}, E17

\bibitem[{{Dickey}(1997)}]{Dickey:1997}
{Dickey}, J.~M. 1997, \apj, 488, 258

\bibitem[{{Foster} \& {MacWilliams}(2006)}]{Foster:MacWilliams:2006}
{Foster}, T., \& {MacWilliams}, J. 2006, \apj, 644, 214

\bibitem[{{Gibson} {et~al.}(2005){Gibson}, {Taylor}, {Higgs}, {Brunt}, \&
  {Dewdney}}]{Gibson:2005}
{Gibson}, S.~J., {Taylor}, A.~R., {Higgs}, L.~A., {Brunt}, C.~M., \& {Dewdney},
  P.~E. 2005, \apj, 626, 195

\bibitem[{{Grant} {et~al.}(2010){Grant}, {Taylor}, {Stil}, {Land ecker},
  {Kothes}, {Ransom}, \& {Scott}}]{Grant:Taylor:2010}
{Grant}, J.~K., {Taylor}, A.~R., {Stil}, J.~M., {et~al.} 2010, \apj, 714, 1689

\bibitem[{{Green} {et~al.}(2012){Green}, {McClure-Griffiths}, {Caswell},
  {Robishaw}, \& {Harvey-Smith}}]{Green:McClure-Griffiths:2014}
{Green}, J.~A., {McClure-Griffiths}, N.~M., {Caswell}, J.~L., {Robishaw}, T.,
  \& {Harvey-Smith}, L. 2012, \mnras, 425, 2530

\bibitem[{{Haffner} {et~al.}(2003){Haffner}, {Reynolds}, {Tufte}, {Madsen},
  {Jaehnig}, \& {Percival}}]{Haffner:Reynolds:2003}
{Haffner}, L.~M., {Reynolds}, R.~J., {Tufte}, S.~L., {et~al.} 2003, \apjs, 149,
  405

\bibitem[{{Kothes} {et~al.}(2004){Kothes}, {Landecker}, \&
  {Wolleben}}]{Kothes:2004}
{Kothes}, R., {Landecker}, T.~L., \& {Wolleben}, M. 2004, \apj, 607, 855

\bibitem[{{Landecker} {et~al.}(2010){Landecker}, {Reich}, {Reid}, {Reich},
  {Wolleben}, {Kothes}, {Uyan{\i}ker}, {Gray}, {Del Rizzo}, {F{\"u}rst},
  {Taylor}, \& {Wielebinski}}]{Landecker:Reich:2010}
{Landecker}, T.~L., {Reich}, W., {Reid}, R.~I., {et~al.} 2010, \aap, 520, A80

\bibitem[{{Liu} {et~al.}(2019){Liu}, {Anderson}, {McIntyre}, {Anish Roshi},
  {Churchwell}, {Minchin}, \& {Terzian}}]{Liu:Anderson:2019}
{Liu}, B., {Anderson}, L.~D., {McIntyre}, T., {et~al.} 2019, \apjs, 240, 14

\bibitem[{{Luisi} {et~al.}(2018){Luisi}, {Anderson}, {Liu}, {Bania}, {Balser},
  {Wenger}, \& {Haffner}}]{Luisi:Anderson:2018}
{Luisi}, M., {Anderson}, L.~D., {Liu}, B., {et~al.} 2018, in American
  Astronomical Society Meeting Abstracts, Vol. 231, American Astronomical
  Society Meeting Abstracts \#231, 230.06

\bibitem[{{O'Sullivan} {et~al.}(2012){O'Sullivan}, {Brown}, {Robishaw},
  {Schnitzeler}, {McClure-Griffiths}, {Feain}, {Taylor}, {Gaensler}, {Land
  ecker}, {Harvey-Smith}, \& {Carretti}}]{OSullivan:Brown:2012}
{O'Sullivan}, S.~P., {Brown}, S., {Robishaw}, T., {et~al.} 2012, \mnras, 421,
  3300

\bibitem[{{Rajabi} {et~al.}(2019){Rajabi}, {Houde}, {Bartkiewicz}, {Olech},
  {Szymczak}, \& {Wolak}}]{Rajabi:Houde:2019}
{Rajabi}, F., {Houde}, M., {Bartkiewicz}, A., {et~al.} 2019, \mnras, 484, 1590

\bibitem[{{Winkel} {et~al.}(2016){Winkel}, {Kerp}, {Fl{\"o}er}, {Kalberla},
  {Ben Bekhti}, {Keller}, \& {Lenz}}]{Winkel16}
{Winkel}, B., {Kerp}, J., {Fl{\"o}er}, L., {et~al.} 2016, \aap, 585, A41

\end{thebibliography}

\end{document}